\documentclass[12pt,a4paper,twoside]{amsart}

\usepackage{amsmath,amsfonts,amsthm}
\usepackage[author-year]{amsrefs}
\usepackage{typearea}
\usepackage{fancyhdr}
\usepackage{float}
\usepackage{graphicx}
\usepackage{algorithm}
\usepackage{algorithmic}
\usepackage[author-year]{amsrefs}

\makeatother

\begin{document}

 {\small\hfill Research Reports MdH/IMa}

{\small\hfill No. 2007-4, ISSN 1404-4978}

\vspace{15mm} \addtolength{\topmargin}{-0.7cm}

\title[]{Efficient Implementation of the AI-REML Iteration\\
for Variance Component QTL Analysis\\}

\author[]{Kateryna Mishchenko}
\address{Department of Mathematics and
Physics, M\"{a}lardalen University, Box 883, SE-721 23
V\"{a}ster{\aa}s, Sweden}
 \email{kateryna.mishchenko@mdh.se}
\author[]{Sverker Holmgren}
\address{Division of Scientific Computing, Department of
Information Technology, Uppsala University, Sweden}
 \email{sverker@it.uu.se}
\author[]{ Lars R\"onneg\aa rd}
\address{Linn{\ae}us Center for Bioinformatics,
Uppsala University, Sweden}
 \email{lars.ronnegard@lcb.uu.se }

\date{\today}

\begin{abstract}Regions in the genome that affect complex traits, quantitative trait
loci (QTL), can be identified using statistical analysis of genetic
and phenotypic data. When restricted maximum-likelihood (REML)
models are used, the mapping procedure is normally computationally
demanding. We develop a new efficient computational scheme for QTL
mapping using variance component analysis and the AI-REML algorithm.
The algorithm uses an exact or approximative low-rank representation
of the identity-by-descent matrix, which combined with the Woodbury
formula for matrix inversion results in that the computations in the
AI-REML iteration body can be performed more efficiently. For cases
where an exact low-rank representation of the IBD matrix is
available a-priori, the improved AI-REML algorithm normally runs
almost twice as fast compared to the standard version. When an exact
low-rank representation is not available, a truncated spectral
decomposition is used to determine a low-rank approximation. We show
that also in this case, the computational efficiency of the AI-REML
scheme can often be significantly improved.
\end{abstract}

\maketitle
\section{Introduction}
Traits that vary continuously are called quantitative. In general,
such traits are affected by an interplay between multiple genetic
factors and the environment. Most medically and economically
important traits in humans, animals and plants are quantitative, and
understanding the genetics behind them is of great importance.

The dissection of quantitative traits is generally performed by
statistical analysis of genetic and phenotypic data for experimental
populations. Such analysis can reveal quantitative trait loci, QTL,
in the genome that affect the trait. In the field of QTL analysis
\cite{LyWa97}, the use of variance component models are commonly
used, see e.g. \cite{Blang01}. Using such models, the statistical
analysis is performed using maximum-likelihood (ML) or restricted
maximum-likelihood (REML) estimators which have the advantage that
they do not use specific assumptions on the design or balance of the
data. These types of schemes are often considered to be
statistically efficient since they utilize all available data. On
the other hand, the corresponding algorithms are also relatively
computationally demanding since non-linear optimization problems
must be solved using an iterative procedure. When searching for the
most likely locations of the QTL, REML optimization problems must be
solved for many genetic locations within an outer global
optimization procedure. Furthermore, if the significance of the
results is established using an experimental procedure like
parametric bootstrap \cite{DaHi97}, hundreds or thousands of QTL
searches must be performed. This implies that the computational
complexity for the numerical methods for solving the REML problems
must be minimal for the QTL mapping to be viable.

During the last two decades, specialized algorithms for variance
component analysis in animal breeding settings have been developed
and implemented in codes like ASREML, DMU and VCE, see \cite{DrDu06}
and references therein. The algorithms used in these codes have in
common that they use a specific Newton-type iteration, the{\em
average information restricted maximum likelihood algorithm},
AI-REML. However, the structure of the problems in traditional
animal breeding applications is different from that of QTL analysis
problems, and the computational schemes should be examined and
possibly modified before they are applied in a QTL analysis setting.
In this short paper we perform this type of investigation and
develop a new efficient computational scheme for QTL mapping using
variance component analysis and the AI-REML algorithm. This type
scheme should then be applied within an efficient and robust
optimization method for maximizing the likelihood in the AI-REML
method, and also within an efficient and robust global optimization
algorithm for the search for the outer optimization problem where
the most likely position of the QTL is determined.

\section{The AI-REML algorithm}
A general linear mixed model is given by
\begin{equation}
y = Xb + Rr + e, \label{LMM}
\end{equation}
where $y$ is a vector of $n$ observations, $X$ is the $n\times
n_{f}$ design matrix for $n_f$ fixed effects, $R$ is the $n\times
n_r$ design matrix for $n_r$ random effects, $b$ is the vector of
$n_{f}$ unknown fixed effects, $r$ is the vector of $n_{r}$ unknown
random effects, and $e$ is a vector of $n$ residuals. In the QTL
analysis setting, we assume that the entries of $e$ are identically
and independently distributed and there is a single observation for
each individual in the pedigree. In this case, the  covariance
matrices are given by $var(e) = I\sigma_e^2$ and
$var(r)=A\sigma_a^2$, where $A$ is referred to as the {\em
identity-by-descent} (IBD) matrix. Using these assumptions, we have
that
\begin{equation}
  var(y) \equiv V = \sigma^{2}_{a} A+ \sigma^{2}_{e}I \equiv
\sigma_1 A +\sigma_2 I\label{Vdef}.
\end{equation}
Here, $\sigma_1\ge 0$ and $\sigma_2>0$ and we assume that the
phenotype follows a normal distribution with $y~MVN(Xb,V)$. At least
two different procedures can be used for computing estimates of $b$
and $\sigma_{1,2}$. In the standard codes mentioned above, the
parameters are computed from the mixed-model equations (MME)
\cite{LyWa97} using a the inverse of $A$. To be able to use this
approach, $A$ has to be positive definite, or it must be modified so
that this property holds. Also, the computations can be performed
efficiently if $A$ is sparse. For the QTL analysis problems, the IBD
matrix $A$ is often only semi-definite and not necessarily sparse.
Here the values of $\sigma_{1,2}$ are given by the solution of the
minimization problem
\begin{eqnarray}
Min\ \ L\, , \\
s.t. \ \ \sigma_{1}\ge 0\nonumber\\
\ \ \ \sigma_{2}> 0
\end{eqnarray}
where $L$ is the log-likelihood for the model (\ref{LMM}),
\begin{equation}
L = -2ln(l) = C + ln(det(V))+ ln(det(X^{T}V^{-1}X))+ yP^{T}y,
\label{LogLikelihood}
\end{equation}
and the projection matrix $P$ is defined by
\begin{equation}
P =V^{-1} - V^{-1}X(X^{T}V^{-1}X)^{-1}X^{T}V^{-1}.\label{P}
\end{equation}
In the original AI-REML algorithm, the minimization problem is
solved using the standard Newton scheme but where the Hessian is
substituted by the {\em average information matrix} $H^{AI}$, whose
entries are given by
\begin{equation}
H^{AI}_{i,j}=y^TP\frac{\partial V}{\partial \sigma_i}P
\frac{\partial V}{\partial \sigma_j}Py\quad,\;i,j=1,2.\label{H}
\end{equation}
The entries of the gradient of $L$ are given by
\begin{equation}
\frac{\partial L}{\partial \sigma_i}=tr\left(\frac{\partial V}
{\partial \sigma_i}P\right)-y^TP\frac{\partial V}{\partial
\sigma_i}Py \quad,\;i=1,2.\label{gradient}
\end{equation}

\section{Efficient implementation of AI-REML optimization}
The main result of this paper is an algorithm that allows for an
efficient implementation of the iteration body in the AI-REML
algorithm for variance component analysis in QTL mapping problems.
In \cite{KaRoHo07}, we examine optimization schemes for the REML
optimization and different approaches for introducing the
constraints for $\sigma_{1,2}$ in the AI-REML scheme. For cases when
the solution to the optimization problem is close to a constraint,
different ad-hoc fixes have earlier been attempted to ensure
convergence for the Newton scheme. In \cite{KaRoHo07}, we show that
by introducing an active-set formulation in the AI-REML optimization
scheme, robustness and good convergence properties were achieved
also for cases when the solution is close to a constraint. When
computing the results below, we use the active-set AI-REML
optimization scheme from \cite{KaRoHo07}.

 From the formulas in the previous section, it is clear that the most
computationally demanding part of the AI-REML algorithm is the
computation of the matrix $P$, and especially the explicit inversion
of the matrix $V = \sigma_{1}A + \sigma_{2}I$, where $A$ is a
constant semi-definite matrix and $\sigma_{1,2}$ are updated in each
iteration. If $V$ is regarded as a general matrix, i.e. no
information about the structure of the problem is used, a standard
algorithm based on Cholesky factorization with complete pivoting is
the only alternative, and ${ O}(n^3)$ arithmetic operations are
required. If $A$ is very sparse, the work can be significantly
reduced by using a sparse Cholesky factorization, possibly combined
with a reordering of the equations. This can be compared to the
approach taken e.g. in \cite{DB04}, where a sparse factorization is
used for $A$ when solving the mixed-model equations. However, as
remarked earlier $A$ is only semi-definite and not very sparse for
the QTL analysis problems and this approach is not an option here.

The IBD matrix $A$ is a function of the location in the genome where
the REML model is applied. The key observation leading to a more
efficient algorithm for inverting $V$ is that the rank of $A$ at a
location with complete genetic information only depends on the size
of the base generation. At such locations, $A$ will be a rank-$k$
matrix where $k\ll n$ in experimental pedigrees with small base
generations. If the genetic information is not fully complete, $A$
can still be approximated by a low-rank matrix. The error in such an
approximation can be made small when the genetic distance to a
location with complete data is small.

Let a symmetric rank-$k$ representation of $A$ be given by
\begin{equation}
A=ZZ^T,\label{low-rank}
\end{equation}
where $Z$ is an $n\times k$ matrix. We now exploit this type of
representation to compute $V^{-1}$.

{\bf Case 1: An exact low-rank representation of $A$ is available
a-priori.} In general, the inverse of a matrix plus a low-rank
update can be computed using the Woodbury formula, see e.g.
\cite{GovL},
\begin{equation}
B^{-1} = (C + S_{1}S_{2}^{T})^{-1} = C^{-1} - C^{-1}S_{1}(I +
S_{2}^{T}C^{-1}S_{1})^{-1} S_{2}^{T}\cdot C^{-1}\label{11}
\end{equation}
Applying (\ref{11}) with $C = I\sigma_{2}$, $S_{1} = Z$ and $S_{2} =
\sigma_{1}Z$, we get the following formula for computing $V^{-1}$,
\begin{equation}
  V^{-1} = I\sigma_{2}^{-1} -
\sigma_{2}^{-1}\widehat{\sigma}Z(I +
\widehat{\sigma}Z^{T}Z)^{-1}Z^{T},\label{18}
\end{equation}
where $\widehat{\sigma} = \sigma_{1}\sigma_{2}^{-1}$. The $k\times
k$ matrix $Z^{T}Z$ can be computed once, before the Newton iteration
is started. Also, the matrix $(I+\widehat{\sigma}Z^TZ)$ is only of
size $k \times k$, and its inverse can be computed using a standard
factorization method in ${O}(k^3)$ arithmetic operations.

{\bf Case 2: No low-rank representation of $A$ is given a-priori.}
If only the matrix $A$ is given, a low-rank
representation/approximation can be computed once before the Newton
iteration is started. Then, the Woodbury formula is again used to
get an efficient computational algorithm for the iterations.
Computing a low-rank approximation of a matrix is a problem with a
long history \cite{EckYou36}, and the standard tool is the truncated
singular value decomposition, see e.g. \cite{GovL,PrTeVe}, which
computes the optimal approximation in Frobenius norm. In our case,
$A$ is symmetric and positive semi-definite, and the SVD is
equivalent to the standard spectral decomposition. Also, the
spectral decomposition can be written as $A=W\Lambda
W^T=(W\sqrt{\Lambda})(W\sqrt{\Lambda})^T$, where $\Lambda={\rm
diag}(\lambda_i)\, ,\; i=1,\ldots ,n$. Hence, we compute a low-rank
representation of $A$ using a truncated spectral decomposition,
\begin{equation}
A = (W_{t}\sqrt{\Lambda_{t}})
(W_{t}\sqrt{\Lambda_{t}})^{T},\label{10}
\end{equation}
where $\Lambda_{t}$ is a diagonal matrix of $k$ eigenvalues and
$W_{t}$ is a rectangular matrix of size $n\times k$ containing the
corresponding orthogonal eigenvectors. Here, $\Lambda_{t}$ and
$W_{t}$ are obtained by eliminating the $n-k$ smallest eigenvalues
and the corresponding eigenvectors from the standard spectral
decomposition. Such idea is used in Genetic Principal Component
Analysis, see e.g. \cite{KiMe04}.

Again using the Woodbury formula (\ref{11}), now with $C =
I\sigma_{2}$, $S_1 = W_{t}\sqrt{\Lambda_{t}}$ and $S_2 =
\sigma_{1}W_{t}\sqrt{\Lambda_{t}}$, we compute  $\tilde V^{-1}$
according to,
\begin{equation}
\tilde V^{-1} = I\sigma_{2}^{-1} - \sigma_{2}^{-1}\widehat{\sigma}
(W_{t}\sqrt{\Lambda_{t}})(I + \widehat{\sigma}
\Lambda_{t})^{-1}(W_{t}\sqrt{\Lambda_{t}})^{T}. \label{12}
\end{equation}

Or in equivalent form,
\begin{equation}
\tilde V^{-1} = I\sigma_{2}^{-1} - \sigma_{2}^{-1}\widehat{\sigma}
W_{t}(\Lambda_{t}^{-1} + \widehat{\sigma}I)^{-1}W_{t}^{T}.
\label{12revised}
\end{equation}

 A similar approach for matrix inversion has been used for
problems in Gaussian process regression with applications in e.g.
machine learning, see \cite{SchTre02,RasWil05}. However, as far as
we know, an approach based on a truncated spectral decomposition and
formula (\ref{12}) has not been been pursued for Newton iterations
for REML-problems before.

When performing the computation described by (\ref{12revised}), the
matrix $W_{t}$ is constant and can be precomputed before the
iteration starts. The matrix $(\Lambda_{t}^{-1} +
\widehat{\sigma}I)$ is diagonal, so its inverse is trivially
computable.

The factorization (\ref{10}) is a rank-$k$ representation of $A$.
This is exact if $A$ has at least $n-k$ zero eigenvalues, and then
$\tilde V^{-1} = V^{-1}$. For a general IBD matrix $A$, more than
$k$ eigenvalues will normally be non-zero, resulting in that $\tilde
V^{-1}$ is an approximation of $V^{-1}$. The accuracy of this
approximation depends on the quality of the truncated factorization
(\ref{10}) for $A$. For the QTL analysis problems, we need to chose
the rank $k$ such that the accuracy of the solution to the REML
problem, i.e. the variance components $\sigma_{1,2}$, is sufficient.

Assume that the eigenvalues are ordered such that $\lambda_1\ge
\lambda_2\ge\,\cdots\,\ge\lambda_n$. Several criteria for choosing
$k$ can be considered. One option is to simply set $k$ to a
predetermined value, either motivated by the known value of the rank
at a fully informative position in the genome or determined as a
given percentage of $n$. Such a choice has the advantage that it is
possible to pre-determine the computational work needed for
computation of $\tilde V^{-1}$ in (\ref{12revised}). Another easily
applied criterion for determining $k$ is given by
\begin{equation}
k = k(\tau) = \{max \ \ i: \lambda_{i}\ge \tau\lambda_1\},
\label{15}
\end{equation}
where the threshold $\tau$ is for example chosen as $\tau = 0.001$.
This means that all eigenvalues of the matrix $A$ that are smaller
than $0.001\lambda_1$ are neglected and the corresponding columns of
$W$ in the spectral decomposition are deleted. In the numerical
experiments presented in the next section, we use the truncation
criterion (\ref{15}).

For computing the truncated spectral decomposition, we currently use
a standard eigenvalue solver, compute all eigenvalues and
eigenvectors, and then ignore the data connected to eigenvalues
smaller than given by the truncation criterion. For problems with
large data matrices where it is known that only a small number of
eigenvalues will be large, an alternative can be to use e.g. the
Lanczos iteration for computing these. Potentially, this can reduce
the arithmetic work for computing the truncated decomposition before
the iteration starts.

Once $V^{-1}$ has been computed using either (\ref{12revised}) or
(\ref{18}), formula (\ref{P}) is used to compute $P$, and the
entries of the Hessian $H^{AI}$ and the gradient of $L$ are computed
using the formulas (\ref{H}) and (\ref{gradient}). The matrices
involved have common terms which are computed once at the beginning
of the current iteration to reduce the total complexity.
\section{Numerical Results}

In this section we present numerical experiment computed using
experimental IBD matrices of size $767\times 767$ drawn from a
chicken population \cite{Ke03}. For these first pilot investigations
of the computational performance we have chosen to implement the
algorithms in Matlab, and we present execution time measurements for
the Matlab scripts. At a later stage, we will implement the most
promising algorithms also in C, which will reduce the execution time
further. In this case, we expect that the relative gain from using
the new algorithms will be larger than indicated below. Matlab's
native implementation of standard inversion is presumably more
efficient than our Matlab scripts for the new algorithms, which are
interpreted at runtime.

\begin{figure}[htbp]
\centering
\includegraphics[width=4in]{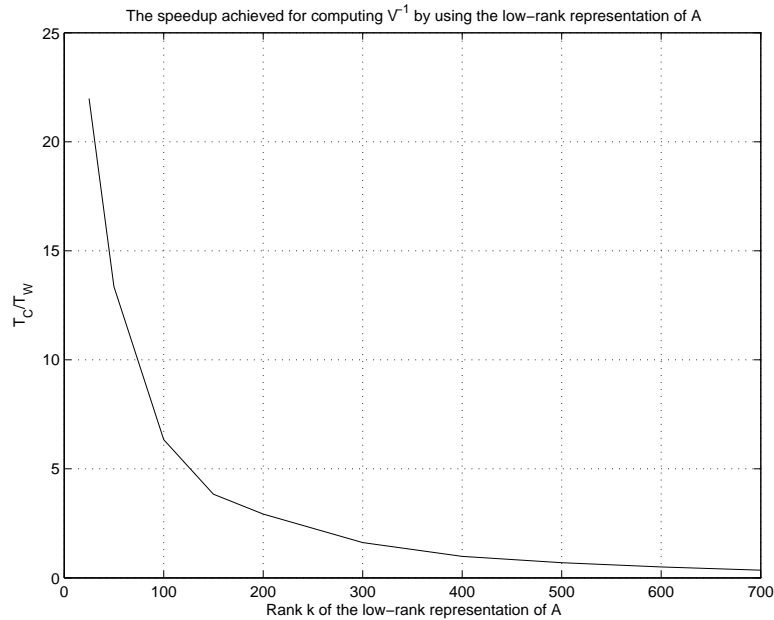}
\caption{The speedup achieved for the computations of $V^{-1}$ in
the AI-REML scheme achieved by using the scheme based on (\ref{18})
instead of standard inversion}\label{f0}
\end{figure}

{\bf Case 1:} We first present numerical results for the case when
the IBD matrices $A$ have exact low-rank representations given
a-priori. Let $T_{C}$ be the measured computational time for the
computation of $V^{-1}$ within the AI-REML scheme using standard
inversion, and let $T_{W}$ be the corresponding time when using the
low-rank representation and formula (\ref{18}). In Figure \ref{f0}
we show $T_{C}/T_{W}$, i.e. the speedup for the computations of
$V^{-1}$, as a function of the rank $k$ of the matrices $A$.

 From Figure \ref{f0}, it is clear that for IBD matrices of size
$767\times 767$, the implementation using the low-rank
representation is faster if the rank $k$ is smaller than
approximately 350. Also, for $k<50$, the speedup is more than one
order of magnitude.

In Table \ref{tab1} we present the total CPU-times for the AI-REML
scheme and the CPU-time for the computations of $V^{-1}$ within this
scheme for different values of $k$ and number of iterations equal to
9.
  The total CPU-times for the
AI-REML are denoted by $T^{tot}_{C}$ and $T^{tot}_{W}$, while the
CPU-times for the computation of $V^{-1}$ are denoted by $T_{C}$ and
$T_{W}$.
\begin{table}[htbp]
\begin{center}
\begin{tabular}{|c|c|c|c|c|}
  \hline
$k$ &$T^{tot}_{C}$& $T_{C}$& $T^{tot}_{W}$&$T_{W}$\\
\hline
700& 10.28 & 5.45 & 20.3 & 15.5 \\
600& 10.28 & 5.45 & 15.7 & 10.9 \\
500& 10.28 & 5.45 & 12.6 & 7.81 \\
400& 10.28 & 5.45 & 10.4 & 5.55 \\
300& 10.28 & 5.45 & 8.20 & 3.37 \\
200& 10.28 & 5.45 & 6.62 & 1.86 \\
150& 10.28 & 5.45 & 6.25 & 1.42 \\
100& 10.28 & 5.45 & 5.69 & 0.86 \\
50& 10.28 & 5.45 & 5.23 & 0.41 \\
25& 10.28 & 5.45 & 5.08 & 0.25 \\
\hline
\end{tabular}
\end{center}
\caption{Total CPU-time for the AI-REML computations and CPU-time
for computing $V^{-1}$ for the schemes using standard inversion and
a low-rank representation}\label{tab1}
\end{table}

 From Table \ref{tab1} it is clear that for the problems studied
here, computing $V^{-1}$ using standard inversion accounts for
approximately half of the computational time in the AI-REML scheme.
For small values of $k$, the time needed for inverting $V^{-1}$
using the low-rank representation can effectively be ignored, and
the improved algorithm runs approximately twice as fast as the
original scheme using standard inversion.

{\bf Case 2:} We now turn to case 2. Here a low-rank representation
of $A$ is not available a-priori, and a truncated eigenvalue
decomposition must be computed before the AI-REML iteration starts.
We present results from computations performed using 18 different
IBD matrices corresponding to genetic locations where a low-rank
representation is not given. The first set of results shows the
computational time $T_{C,TW}$ for computing $V^{-1}$ within the
AI-REML scheme. In this case we compare standard inversion (C) to
the scheme given by (\ref{12}), based on truncated spectral
decomposition (TW).

In Table \ref{tab2}, we show the CPU-time and the number of AI-REML
iterations for the two schemes. For the scheme based on the
truncated spectral decomposition, the table also shows the rank $k$
as determined by the truncation criterion (\ref{15}) and the
relative errors (in \%) for the variance components
($e_{\sigma_{1,2}}$) arising from using a low-rank approximation of
$A$.
\begin{table}[hbtp]
\begin{center}
\begin{tabular}{|c|c|c|c|c|c|c|c|c|c|c|c|c|}
  \hline
&\multicolumn{2}{|c|}{C}&\multicolumn{10}{|c|}{TW}\\
\cline{2-13} A & & & \multicolumn{5}{|c|}{$\tau$ = 0.001}&
     \multicolumn{5}{|c|}{$\tau$ =0.005}\\
\cline{4-13} &$T_{C}$&\# it&$T_{TW}$&k & $ \# it$ & $e_{\sigma_{1}}$
[\%] & $e_{\sigma_{2}}$ [\%] & $T_{TW}$& $k$ & \# it &
$e_{\sigma_{1}}$ [\%] &
$e_{\sigma_{2}}$ [\%]\\
\hline
  $A_{1}$& 4.18 & 7 & 2.20 & 294 & 7 & 0.82 & 1.67
& 0.531 & 82 & 7 &
6.7 & 4.7\\
  $A_{2}$& 4.16 & 7 & 1.12 & 166 & 7 & 0.03 & 0.377
& 0.203 & 37 & 6 &
7.3 & 1.8\\
  $A_{3}$& 3.50 & 6 & 1.25 & 169 & 7 & 0.11 & 1.19
& 0.307 & 35 & 8 &
12.7 & 3.0\\
  $A_{4}$& 3.48 & 6 & 1.87 & 280 & 6 & 1.14 & 1.83
& 1.05 & 71 & 15 &
40.8 & 7.5\\
  $A_{5}$& 2.81 & 5 & 1.81 & 291 & 6 & 0.94 & 1.56
& 0.920 & 82 & 11 &
30.4 & 5.4\\
  $A_{6}$& 4.89 & 8 & 2.60 & 294 & 8 & 0.64 & 0.76
& 0.438 & 81 & 6 &
8.1 & 3.1\\
  $A_{7}$& 2.88 & 5 & 1.66 & 286 & 5 & 0.84 & 0.53
& 0.313 & 70 & 5 &
5.9 & 2.1\\
  $A_{8}$& 4.93 & 8 & 1.22 & 163 & 8 & 0.44 & 0.18
& 0.314 & 43 & 7 &
3.9 & 0.64\\
  $A_{9}$& 8.36 & 13 & 0.671 & 54 & 12 & 0.04 &
0.03 & 0.328 & 15 & 13 &
4.5& 0.09\\
  $A_{10}$& 15.3 & 23 & 1.22 & 50 & 23 & 0.07& 0.02
& 0.527 & 10 & 23 &
0.05 & 0.05\\
  $A_{11}$& 1.98 & 4 & 0.140 & 46 & 4 & - & 0 &
0.093 & 13 & 4 & - & 0\\
  $A_{12}$& 2.00 & 4 & 0.344 & 71 & 5 & - & 0 &
0.108 & 14 & 5 & - & 0\\
  $A_{13}$& 5,45 & 9 & 0.467 & 57 & 9 & 0.004& 0
& 0.173 & 12 & 9 & 0.02
& 0.02\\
  $A_{14}$& 2.86 & 5 & 0.873 & 192 & 5 & 0.53 &
0.26 & 0.327 & 51 & 7 &
16.9 & 1.6\\
  $A_{15}$& 2.77 & 5 & 0.884 & 198 & 5 & 0.36 &
0.41 & 0.250 & 51 & 5 &
3.1 & 1.1\\
  $A_{16}$& 2.83 & 5 & 0.749 & 173 & 5 & 0.19 &
0.39 & 0.327 & 44 & 8 &
22.2 & 2.0\\
  $A_{17}$& 2.80 & 5 & 1.28 & 270 & 5 & 0.96 & 0.95
& 0.311 & 60 & 6 &
6.3 & 3.2\\
  $A_{18}$& 4.23 & 7 & 2.62 & 303 & 8 & 0.69 & 0.91
& 0.485 & 86 & 6 &
8.0 & 3.7\\
\hline
\end{tabular}
\end{center}
\caption{CPU-time for computing $V^{-1}$ within the AI-REML
iteration,
  and errors arising from exploiting a low-rank approximation of
$A$}\label{tab2}
\end{table}
For the matrices $A_{11}$ and $A_{12}$, the variance component
$\sigma_1$ is zero, and it is not possible to compute the relative
error. For these matrices, the absolute errors in $\sigma_1$ is very
small for both values of $\tau$. From the results in Table
\ref{tab2}, it is clear that the accuracy of the variance components
does not only depend on the value of the parameter $\tau$, but also
on the properties of the matrix $A$. When an approximation of
$V^{-1}$ is employed in the computations, the optimization landscape
is affected in different ways for different IBD matrices $A$,
resulting in different behavior for the optimization scheme and
different errors in the location of the optima. However, for the
problems studied here, choosing $\tau=0.001$ results in that the
variance components are determined with sufficient accuracy for all
the IBD matrices examined.

In the next set of experiments we study the speedup and error
introduced by using an approximate inverse of $V$ in some more
detail for the two IBD matrices $A_1$ and $A_{13}$. As can be seen
from Table \ref{tab2}, $A_1$ is harder to approximate with a low
rank matrix than $A_{13}$. In Tables \ref{tab3} and \ref{tab4}, we
vary the value of the parameter $\tau$ in the truncation criterion
(\ref{15}) and show the resulting rank $k$ of the approximation of
$A$, the number of iterations in the AI-REML optimization scheme,
the speedup $T_{C}/T_{TW}$, and the relative errors in the variance
components.

\begin{table}[htbp]
\begin{center}
\begin{tabular}{|c|c|c|c|c|c|c|c|c|c|c|c|c|}
  \hline
$\tau$&$k$ &\# it&$T_{C}/T_{TW}$& $e_{\sigma_{1}}$ [\%] &
$e_{\sigma_{2}}$ [\%] \\
\hline
$10^{-7}$ & 699 & 7 & $0.46$ &  0.000 & 0.000 \\
$10^{-6}$ & 662 & 7 & $0.51$ &  0.002 & 0.001 \\
$10^{-5}$ & 556 & 7 & $0.71$ &  0.070 & 0.054 \\
$10^{-4}$ & 294 & 7 & $1.97$ &  0.82 & 1.67 \\
$5\cdot 10^{-4}$ & 82 & 7 & $7.68$ & 6.75 & 4.69 \\
$10^{-3}$ & 29 & 36 & $3.28$ & 95.2 & 6.70 \\
$5\cdot 10^{-3}$ & - & no conv. & - & - & - \\
\hline
\end{tabular}
\end{center}
\caption{Results for the IBD matrix $A_1$ for different values of
the truncation parameter $\tau$}\label{tab3}
\end{table}

\begin{table}[htbp]
\begin{center}
\begin{tabular}{|c|c|c|c|c|c|c|c|c|c|c|c|c|}
  \hline
$\tau$&$k$ &\# it&$T_{C}/T_{TW}$& $e_{\sigma_{1}}$ [\%] &
$e_{\sigma_{2}}$ [\%] \\
\hline
$10^{-7}$ & 134 & 9 & $6.41$ & 0.000  & 0.000 \\
$10^{-6}$ & 134 & 9 & $6.50$ & 0.000  & 0.000 \\
$10^{-5}$ & 123 & 9 & $7.12$ & 0.000  & 0.000 \\
$10^{-4}$ & 57 & 9 & $15.0$ & 0.004 & 0.006 \\
$5\cdot 10^{-4}$ & 12 & 9 & $34.6$ & 0.009 & 0.025\\
$10^{-3}$ & 9 & 9 & 37.3 & 0.024 & 0.025 \\
$5\cdot 10^{-3}$ & 7 & 9 & 44.7 & 0.109 & 0.027\\
$10^{-2}$ & 5 & 9 & 55.7 & 2.3 & 0.046 \\
$5\cdot 10^{-2}$ & - & no conv. & - & - & - \\
\hline
\end{tabular}
\end{center}
\caption{Results for the IBD matrix $A_{13}$ for different values of
the truncation parameter $\tau$}\label{tab4}
\end{table}

\begin{figure}[htbp]
\centering
\includegraphics[width=4.5in]{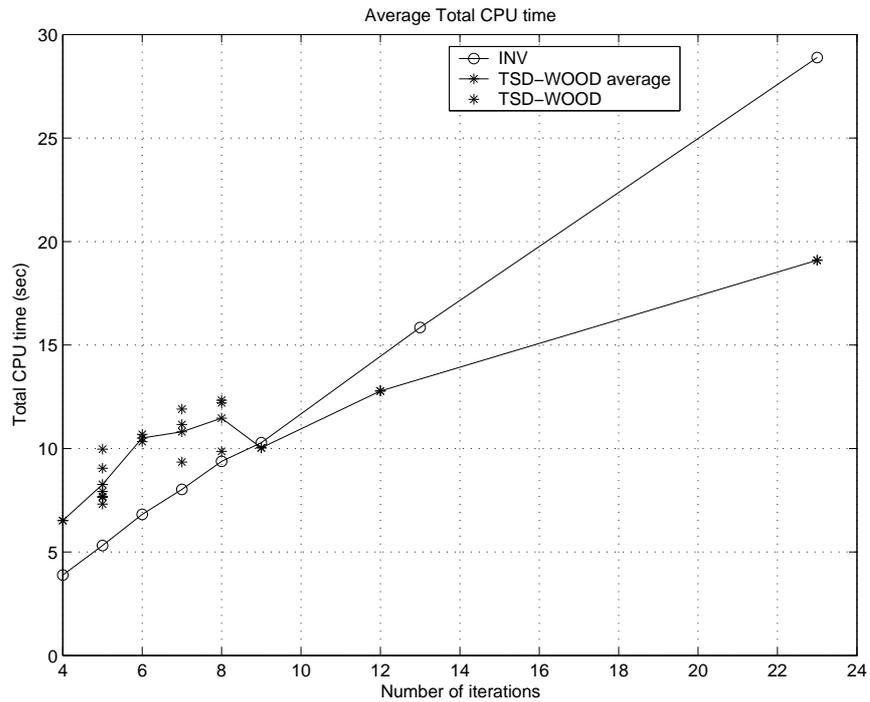}
\caption{Total CPU-times for solving the AI-REML problems using the
scheme based on standard inversion of $V$ and the new scheme using a
truncated spectral decomposition}\label{f1}
\end{figure}

In Figure \ref{f1} we finally present total timings for solving the
AI-REML problems using the new scheme exploiting a truncated
spectral decomposition, including the time needed for computing the
truncated factorization prior to the AI-REML iterations. We also
compare these timings to the corresponding results for the standard
scheme using direct inversion. In the figure, the timings for all
matrices $A_1 - A_{18}$ are shown as a function of the number of
AI-REML iterations required. When several matrices result in the
same number of iterations, the cpu time for each matrix and the
average result are shown.

Theoretically, the average cpu time for the scheme exploiting the
truncated spectral decomposition should be a straight line.
Deviations are due to the inconsistence of time measurement in
Matlab.

 From Figure \ref{f1}, we see that the standard scheme using direct
inversion is faster when number of iterations is small, i.e. less
than $9$ iterations. The reason for this is that the spectral
decomposition of the matrix A is relatively costly, and it must be
amortized over a number of iterations before the faster iterations
begin to pay of. When the number of iterations required is
significantly larger than $9$, which is often the case for
real-world problems, the new algorithm is significantly faster also
when no low-rank representations of the IBD matrices are not
available a priori.

\section{Conclusions}
In this paper we present a family of algorithms that allow for an
efficient implementation of the iteration body in the AI-REML
algorithm for variance component analysis in QTL mapping problems.
Combined with the improved optimization scheme in \cite{KaRoHo07},
the new algorithms form a basis for an efficient and robust AI-REML
scheme for evaluating variance component QTL models.

The most costly operation in the AI-REML iteration body is the
explicit inversion of the matrix $V = \sigma_{1}A + \sigma_{2}I$,
where the IBD matrix $A$ is constant and positive semi-definite, and
$\sigma_{1}\ge 0$ and $\sigma_2>0$ are updated in each iteration.
The key information enabling the introduction of improved algorithms
is that the rank of $A$ at a location in the genome with complete
genetic information only depends on the size of the base generation.
At such locations, $A$ will be a rank-$k$ matrix where $k\ll n$, and
by exploiting the Woodbury formula the inverse of $V$ can be
computed more efficiently than by using a standard algorithm based
on Cholesky factorization. If the genetic information is not fully
complete, a general IBD matrix $A$ can still be approximated by a
low-rank matrix and the error in such an approximation can be made
small when the genetic distance to a location with complete data is
small. More importantly, there might be a possibility for setting up
a low-rank representation of the matrix $A$ also at genetic
locations where the information is not complete. This is a topic of
current investigation \cite{RC07}.

We present results for IBD matrices $A$ from a real data set for two
different settings; Firstly, we show that if a low-rank
representation of $A$ is available a-priori, the inversion of $V$
using the new algorithms is performed faster than using a standard
algorithm if the rank $k$ is smaller than approximately 350. Also,
for $k<50$, the speedup is more than one order of magnitude.

Then we also show that, even if a low-rank representation is not
directly available and a low-rank approximation of $A$ needs to be
computed before the AI-REML iterations, significant speedup of the
variance component model computations can still be achieved. For QTL
mapping problems, the efficiency of our new method will increase
when the ratio between the total pedigree size and base generation
size increases, the density and informativeness of markers
increases. Hence, the relative efficacy of the method will
continuously increase in the future with deeper pedigrees and more
markers. Also, we are currently developing a scheme for directly
constructing a low-rank representation of the IBD matrix $A$ also
between markers, which will result in that the eigenvalue
factorization prior to the AI-REML iterations is not needed any
more.
\bibliographystyle{plain}
\bibliography{ref_new}

\end{document}